\documentclass[twocolumn,letterpaper,showpacs,aps,prd,superscriptaddress]{revtex4}

\usepackage{graphicx}
\usepackage{dcolumn}
\usepackage{epsfig}
\usepackage{psfrag}
\usepackage{amsmath}
\usepackage{hyperref}
\usepackage{relsize}
\RequirePackage{xspace}

\newcommand{\BaBarYear}    {09}
\newcommand{\BaBarNumber}  {030}

\newcommand{\SLACPubNumber} {13837}

 \newcommand{\BaBarType}      {PUB}  

\def\babar{\mbox{\slshape B\kern-0.1em{\smaller A}\kern-0.1em 
    B\kern-0.1em{\smaller A\kern-0.2em R}}}
\mathchardef\Upsilon="7107
\def\Y#1S{\ensuremath{\Upsilon{(#1S)}}\xspace}
\def\BR         {{\ensuremath{\cal B}\xspace}}
\newcommand{\gev}{\ensuremath{\mathrm{\,Ge\kern -0.1em V}}\xspace}
\newcommand{\mev}{\ensuremath{\mathrm{\,Me\kern -0.1em V}}\xspace}
\newcommand{\gevc}{\ensuremath{{\mathrm{\,Ge\kern -0.1em V\!/}c}}\xspace}
\newcommand{\mevc}{\ensuremath{{\mathrm{\,Me\kern -0.1em V\!/}c}}\xspace}
\newcommand{\gevcc}{\ensuremath{{\mathrm{\,Ge\kern -0.1em V\!/}c^2}}\xspace}
\newcommand{\mevcc}{\ensuremath{{\mathrm{\,Me\kern -0.1em V\!/}c^2}}\xspace}
\def\pep2{PEP-II}
\def\invfb   {\ensuremath{\mbox{\,fb}^{-1}}\xspace}
\def\mum  {\ensuremath{{\,\mu\rm m}}\xspace}
\def\xp  {\ensuremath{{\rm x_p}}}
\newcommand{\bef} {\begin{figure}}
\newcommand{\enf}[2]  {\caption{\label{#1}#2}\end{figure}}

\newcommand{\bet}[3]{
  \begin{table}
      \caption{\label{#2}#3}
      \begin{ruledtabular}
      \begin{tabular}{#1}
}

\newcommand{\ent}{
      \end{tabular}
      \end{ruledtabular}
  \end{table}
}

\long\def\inst#1{\par\nobreak\kern 4pt\nobreak
    {\it #1}\par\vskip 10pt plus 3pt minus 3pt}

\begin{document}
\begin{flushleft}
\babar-\BaBarType-\BaBarYear/\BaBarNumber \\
SLAC-PUB-\SLACPubNumber \\
Phys.\ Rev.\ D 81, 011102(R) (2010) \\
\end{flushleft}

\title{{\boldmath Observation of inclusive $D^{*\pm}$ production in the decay of \Y1S}
}


\begin{abstract}
We present a study of the inclusive $D^{*\pm}$ production in the decay
of $\Upsilon(1S)$ using $(98.6 \pm 0.9)\times 10^6$ $\Upsilon(2S)$
mesons collected with the 
\mbox{\slshape B\kern-0.1em{\smaller A}\kern-0.1em
    B\kern-0.1em{\smaller A\kern-0.2em R}} 
detector at the $\Upsilon(2S)$
resonance. Using the decay chain $\Upsilon(2S)\to \pi^+ \pi^- \Upsilon(1S),\;
\Upsilon(1S) \to D^{*\pm} X$, where $X$ is unobserved, we measure the
branching fraction $\BR[\Upsilon(1S)\to D^{*\pm} X] = (2.52 \pm
0.13({\rm stat}) \pm 0.15({\rm syst}))\%$ and the $D^{*\pm}$ momentum
distribution in the rest frame of the $\Upsilon(1S)$. We find evidence for an
excess of $D^{*\pm}$ production over the expected rate from the
virtual photon annihilation process $\Upsilon(1S)\to\gamma^*\to c\bar c\to
D^{*\pm} X$.

\end{abstract}

%
\author{B.~Aubert}
\author{Y.~Karyotakis}
\author{J.~P.~Lees}
\author{V.~Poireau}
\author{E.~Prencipe}
\author{X.~Prudent}
\author{V.~Tisserand}
\affiliation{Laboratoire d'Annecy-le-Vieux de Physique des Particules (LAPP), Universit\'e de Savoie, CNRS/IN2P3,  F-74941 Annecy-Le-Vieux, France}
\author{J.~Garra~Tico}
\author{E.~Grauges}
\affiliation{Universitat de Barcelona, Facultat de Fisica, Departament ECM, E-08028 Barcelona, Spain }
\author{M.~Martinelli$^{ab}$}
\author{A.~Palano$^{ab}$ }
\author{M.~Pappagallo$^{ab}$ }
\affiliation{INFN Sezione di Bari$^{a}$; Dipartimento di Fisica, Universit\`a di Bari$^{b}$, I-70126 Bari, Italy }
% ...
\collaboration{The \babar\ Collaboration}
\noaffiliation

\maketitle

\section{Introduction}
Bound states of heavy quarks provide a powerful testing ground for
quantum chromodynamics (QCD). Experimental studies of charmonium and
bottomonium spectroscopy have helped uncover some of the key aspects
of the quarkonium
potential~\cite{quarkonia_spectroscopy1,quarkonia_spectroscopy2}.
Studies of the decays of quarkonia and of their decay products can also
reveal important information on QCD
processes~\cite{quarkonia_transitions}. The hadronic decays of the
narrow quarkonia, states which are below the threshold for open flavor
production, are dominated by couplings to gluons and the fragmentation
process into light hadrons. The decay properties of charmonia, which
have a relatively low multiplicity particle content, have been
extensively studied~\cite{qwg}.  However, little is known about the
final state contents of bottomonia. In particular, scarcely any
experimental information exists on the decays of bottomonium to open
charm. The CLEO Collaboration has observed~\cite{cleo_chib} charm
production in the decays of the $\chi_b$ states with branching fractions
of the order of 10\%.  The ARGUS Collaboration
searched~\cite{argusy1s} for the decay $\Upsilon(1S)\to D^{*\pm} X$
and set a limit on its branching fraction of $\BR<1.9\%$ at 90\%
confidence level.

In this article, we report a study of the inclusive process
$\Upsilon(1S)\to D^{*\pm} X$, yielding the decay branching fraction and
the $D^{*\pm}$ momentum spectrum in the $\Y1S$ rest frame, using data
recorded by the \babar{} Collaboration at the $\Upsilon(2S)$
resonance. The decay $\Upsilon(1S)\to D^{*\pm} X$ can proceed through
the QED virtual photon annihilation process,
$\Upsilon(1S)\to\gamma^*\to c\bar c$, followed by the hadronization of
the $c \bar c$ system. The expected decay rate and the $D^{*\pm}$
momentum spectrum from this process can be accurately estimated from
the measured properties of the $\Y1S$ decays and the charm
fragmentation function measured at the center-of-mass energy
$\sqrt{s}\sim 10~\gev$.
Other QCD processes such as the splitting of a virtual
gluon~\cite{singlet1,singlet2,singlet3} or the
annihilation of the $b \bar b$ system in an octet state~\cite{octet},
have also been suggested as major contributors to this decay channel.
Measurements of the $D^{*\pm}$ yield and of its momentum spectrum can
help test the predictions of the proposed QCD mechanisms, and possibly
reveal the presence of new physics processes with exotic couplings to
heavy quarks~\cite{ps_higgs,little_higgs}.

\section{The \babar{} detector}
The results presented in this work are based on data collected at
center-of-mass energy corresponding to the mass of $\Y2S$ resonance
with the \babar{} detector at the \pep2{} asymmetric energy $e^+e^-$
storage ring operating at the SLAC National Accelerator
Laboratory. The data consist of 14.4 \invfb of integrated luminosity,
corresponding to $98.6 \pm 0.9 $ million \Y2S mesons produced. The
study of \Y1S decays is performed by reconstructing the decay chain
$\Y2S \to \pi^+\pi^- \Y1S$, which yields approximately $17.8$ million
$\Y1S$ decays.  An additional off-resonance data sample corresponding
to $44.5$ \invfb collected at $\sqrt{s}$ about $40 \mev$ below the
\Y4S resonance is used to study the background. A
GEANT4-based~\cite{geant4} simulation of the detector is used to
determine the properties of the signal and to study the background
sources.

A detailed description of the \babar{} detector can be found
elsewhere~\cite{babarnim}. The tracking system is composed of a 5
layer silicon vertex tracker (SVT) and a 40 layer drift chamber (DCH)
in a $1.5\,{\rm T}$ magnetic field. The SVT provides a precise
determination of the track impact parameters and angles near the
interaction point (IP) with $15\mum$ spatial resolution at normal
incidence at a radius of 3.2 cm, and is capable of stand-alone
tracking for low momentum particles down to $50\mevc$ of transverse
momentum $p_t$.  The DCH, together with the SVT, provides a precise
measurement of the momenta and azimuthal angles of charged particles
with a resolution $\sigma_{p_t}/p_{t}=(0.13~p_t \oplus 0.45)\%$, where
$p_t$ is in units of \gevc.  Charged hadron identification
is achieved through measurements of the specific ionization energy
loss in the SVT and DCH, and of the Cherenkov angle from a detector of
internally reflected Cherenkov light (DIRC).  A CsI(Tl)
electromagnetic calorimeter (EMC) provides photon detection, electron
identification, and $\pi^0$, $\eta$ and $K^0_L$ reconstruction.
Finally, the instrumented flux return (IFR) of the magnet allows
discrimination of muons from pions and detection of neutral kaons.

\section{Candidate Reconstruction and Selection}
The $\Y2S \to \pi^+\pi^- \Y1S$ candidates are identified by forming
pairs of oppositely charged tracks whose recoil mass is consistent
with the mass of the \Y1S resonance, when the tracks are interpreted
as pions. The recoil mass $M_{\rm recoil}$ is computed using
\begin{equation}
M_{\rm recoil} \equiv \sqrt{(P_{e^+e^-} - P_{\pi\pi})^2}
\end{equation}
where $P_{e^+e^-}$ is the known 4-momentum of the $e^+e^-$ system and
$P_{\pi\pi}$ is the reconstructed 4-momentum of the $\pi^+\pi^-$ pair.
The pion tracks are required to have energy losses and Cherenkov angles
consistent with the pion hypothesis. The track pair is fitted to a
common vertex and the probability of the vertex fit is required to be
greater than 1\%.  The measured di-pion mass distribution
peaks near $0.52$\gevcc for
signal events~\cite{cleopipi}, whereas background events are
approximately uniformly distributed in the kinematically allowed mass
interval $[0.28,0.56]\gevcc$. Requiring the mass of the pion pair to
be greater than $0.4\gevcc$ retains 96\% of the signal candidates
while rejecting approximately 1/3 of the background events.
Figure~\ref{fig:mrecoil} shows the recoil mass distribution for the
event sample passing the above selection criteria; a signal region
consisting of two standard deviations around the \Y1S mass is
highlighted (cross hatching), as well as two sideband regions used
for background studies, the lower ($[9432.1,9444.3]\mevcc$) and upper
($[9477.7,9490.0]\mevcc$) sidebands (diagonal shading).  These events
form the full event-set used in the measurement of the $D^{*\pm}$
yield.

We reconstruct $D^{*\pm}$ candidates using the decay chain $D^{*+} \to
D^0 \pi^+$, $D^0 \to K^-\pi^+$.  A kaon candidate and an oppositely
charged pion candidate are combined to form the $D^0$ candidate. The
identification efficiency for kaons (pions) is about 98\% (93\%); the
misidentification rate of kaons (pions) as pions (kaons) is about 5\%
(15\%).  The identification performance is obtained from a control
sample of inclusive $D^{*+}\to D^0\pi^+,D^0\to K^-\pi^+$.  The kaon
and pion tracks are geometrically constrained to originate from a
common vertex and the probability of the vertex fit is required to be
greater than 1\%.  The mass of the $D^0$ candidate is required to be
within $75\mevcc$ of the nominal $D^0$ mass, which corresponds to
about 18 times the experimental resolution on the $D^0$ candidate
mass. This large mass interval is necessary for the subtraction of the
combinatorial background.
The $D^0$ candidate is finally combined with a soft pion with its
charge opposite to that of the kaon candidate to form a $D^{*+}$
candidate. The mass difference between the $D^{*+}$ and the $D^0$
($\Delta m$) is required to be in the interval
$[143.20,147.64]\mevcc$, which corresponds to approximately six times
the experimental resolution.  The soft pion and the $D^0$ candidates
are fitted to a common vertex constrained to originate from the
interaction region. The probability of the $D^{*\pm}$ vertex fit is
required to be greater than 1\%. For events with multiple candidates,
the candidate with the best combined vertex fit $\chi^2$, defined as
the sum of the $\chi^2$ values from the vertex fits described above,
is kept. The multiplicity of the reconstructed candidates in simulated
signal MC events is 1.2, after the final selection. 74\% of these
candidates are correctly matched to a signal candidate. The best
candidate algorithm retains 90\% of the correctly matched candidates
and 68\% of the ones not correctly matched.

\section{Signal Extraction}

The sample of $D^{*\pm}$ candidates is studied in intervals of the
scaled momentum \xp, defined as:
\begin{equation} 
 \xp = \frac{p_{D^{*\pm}}}{p_{\rm max}}
\label{eq:xp}
\end{equation}
where $p_{D^{*\pm}}$ is the $D^{*\pm}$ momentum in the rest frame of
the $\Y1S$, $p_{\rm max} = \sqrt{E_{\rm max}^2-m_{D^{*+}}^2}$, $E_{\rm max} =
m_{\Y1S}/2$ and $m_{D^{*+}}$ is the world average of the $D^{*+}$ mass~\cite{pdg08}.

The sample is divided into \xp{} intervals of 0.05 width in the range
$[0.1,1.0]$; the region $\xp<0.1$, which is dominated by combinatorial
background, is excluded.  

The invariant mass distribution of the $D^0$ candidates in each \xp{}
interval is used to determine the $D^{*\pm}$ yield from $\Y1S \to
D^{*\pm} X$.  The $D^0$ mass distribution is obtained from the
$K^-\pi^+$ candidates mass distribution by two background
subtractions.  Combinatorial backgrounds, events that are not $\Y2S
\to \pi^+ \pi^- \Y1S$ decays, are removed by subtracting the lower and
upper sidebands of the $\pi^+ \pi^-$ recoil mass.  The $K^-\pi^+$
invariant mass distribution from the sidebands is rescaled to the
expected number of background events in the signal region to determine
the $K^-\pi^+$ mass distribution from the combinatorial background
component under the \Y1S peak.  In addition, the $K^-\pi^+$ mass
distribution for ``wrong-sign'' $D^0(\to K^-\pi^+)\pi^-$ combinations
(where the soft pion has the same charge as that of the kaon
candidate) is used to subtract the $D^*$ combinatoric background
including a possible peaking backgrounds from $D^0(\to K^-\pi^+)\pi^+$
combinations, involving a true $D^0$ decay and a random soft
pion. This method leads to a small over-subtraction of signal events
due to doubly Cabibbo suppressed (DCS) $D^0$ decays reconstructed as
wrong-sign combinations. This is accounted for in the final estimation
of the branching fraction.  The background subtracted invariant mass
distribution of $D^0$ candidates in the full \xp{} range is shown in
Figure~\ref{fig:md0sbsub}.

Finally the invariant mass distribution of the $D^0$ candidates in
each \xp{} interval is fitted to a probability density function
(p.d.f.)  using a minimum $\chi^2$ estimator.  The fitted p.d.f.,
$P(m)$, is the sum of a signal p.d.f., $P_{\rm sig}(m)$, and a p.d.f. which
accounts for unsubtracted backgrounds, $P_{\rm bkg}(m)$,
\begin{equation}
  P(m) = n_{\rm sig}\times P_{\rm sig}(m) + n_{\rm bkg}\times P_{\rm bkg}(m)
\end{equation}
where $n_{\rm sig}$ and $n_{\rm bkg}$ are the number of signal and background
events in the fitted region.  The fit region corresponds to the $D^0$
mass range in which we accept signal candidates
[$m_{D^0}-75\mevcc,m_{D^0}+75\mevcc]$. The signal p.d.f. is the sum of
two Gaussian functions with the same mean:
\begin{equation}
P_{\rm sig}(m;f,\mu,\sigma_1,\sigma_2)  =  f G(m;\mu,\sigma_1) + (1-f) G(m;\mu,\sigma_2) 
\end{equation}
The background p.d.f. is a linear function:
\begin{equation}
P_{\rm bkg}(m;\mu,p_1)  =  1/w + p_1 (m-\mu)
\end{equation}
where $w$ is the fit range.  The parameters of the signal p.d.f.,
$\sigma_1$, $\sigma_2$ and $f$ are determined from a fit to the
corresponding distribution from Monte Carlo (MC) simulation.  However,
the mean of the $D^0$ mass, $\mu$, is fixed to the value determined
from a fit to the $D^0$ mass distribution in the full \xp{} interval.

\bef \includegraphics[width=0.49\textwidth]{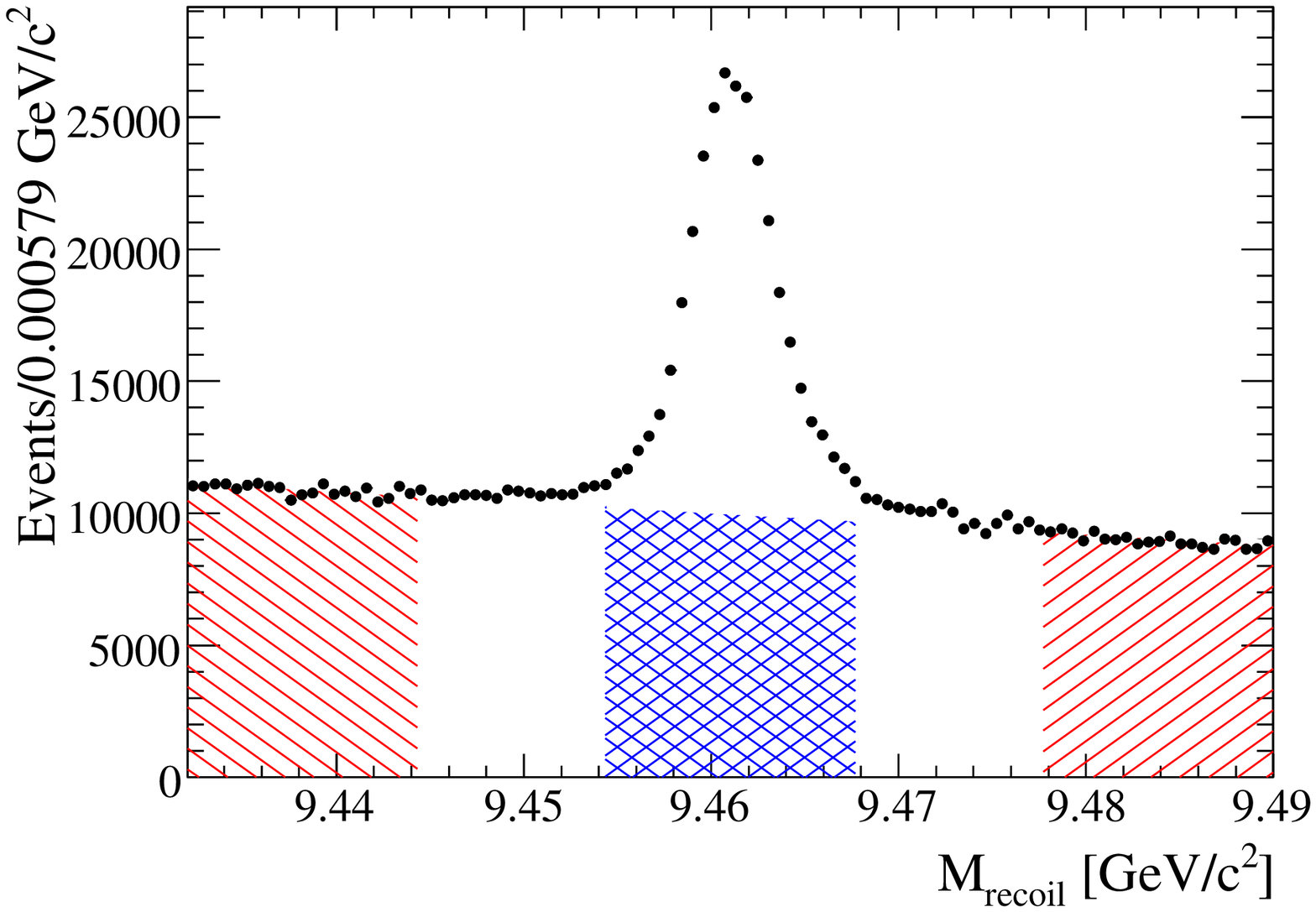}
\enf{fig:mrecoil}{Distribution of the recoil mass, $M_{\rm recoil}$,
  for the selected $\Y2S\to \pi^+\pi^-\Y1S$ events. The cross
  hatching shows the signal region, and the
  lower and upper sideband regions are indicated by the diagonal shading.}

\bef \includegraphics[width=0.49\textwidth]{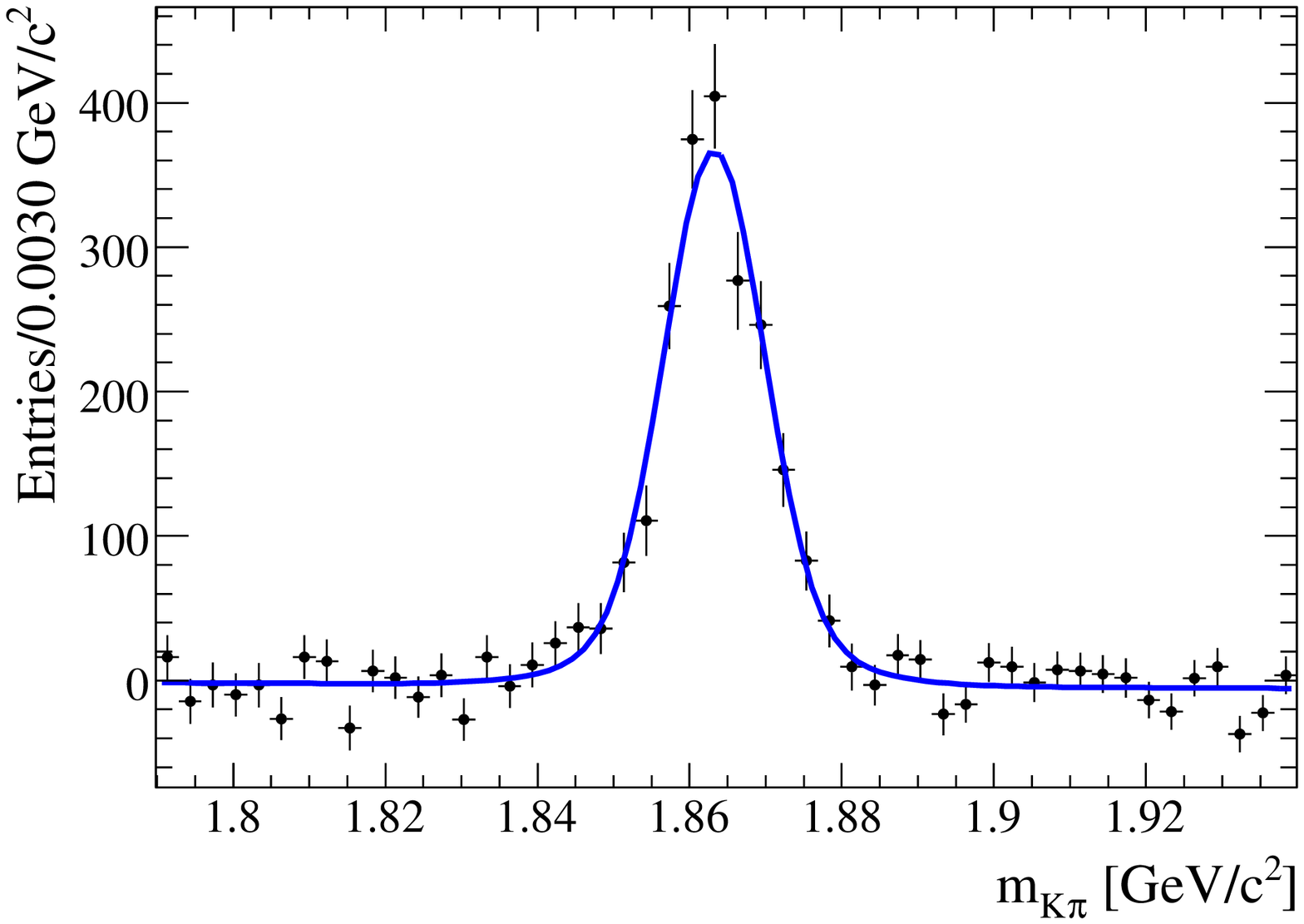}
\enf{fig:md0sbsub}{Distribution of the $D^0$ invariant mass for the
  complete [0.1;1.0] \xp{} range after subtraction of the combinatoric
  background and wrong-sign combinations. The solid line represents the fit
  to the data of the p.d.f. described in the text. }

The event selection efficiency is determined using a simulation study
of the signal and background processes.  Signal events are obtained by
generating the transition $\Y2S \to \pi^+ \pi^- \Y1S$ according to the
decay model determined by CLEO~\cite{cleopipi}, followed by the decay
$\Y1S \to c \bar c$ and the hadronization of the $c \bar c$ pair via
JETSET~\cite{jetset}. Signal events are required to contain at least
one $D^{*\pm}$ meson after the hadronization process. The small
fraction of events (0.4\%) containing both $D^{*+}$ and $D^{*-}$
decays is accounted for by normalizing the efficiency to the number of
signal decays generated. The selection efficiency as a function of
\xp{}, $\epsilon(\xp)$, is shown in Figure~\ref{fig:eff}. The
dependence on \xp{} is mainly due to the reconstruction efficiency of
the slow pion from the $D^{*\pm}$ decay.  The average reconstruction
efficiency in data depends on the measured \xp{} distribution and can
be estimated from the relation $ <\epsilon_{\rm data}> =
\frac{\Sigma_\xp n_{\rm sig}(\xp)}{\Sigma_\xp n_{\rm
    sig}(\xp)/\epsilon(\xp)} = (17.7\pm 0.3)\%$ where the error is
statistical only. The ratio of the $\chi^2$ to the number of degrees
of freedom for the individual fits ranges from 0.5 to 2.5, with 16
degrees of freedom.

\bef \includegraphics[width=0.49\textwidth]{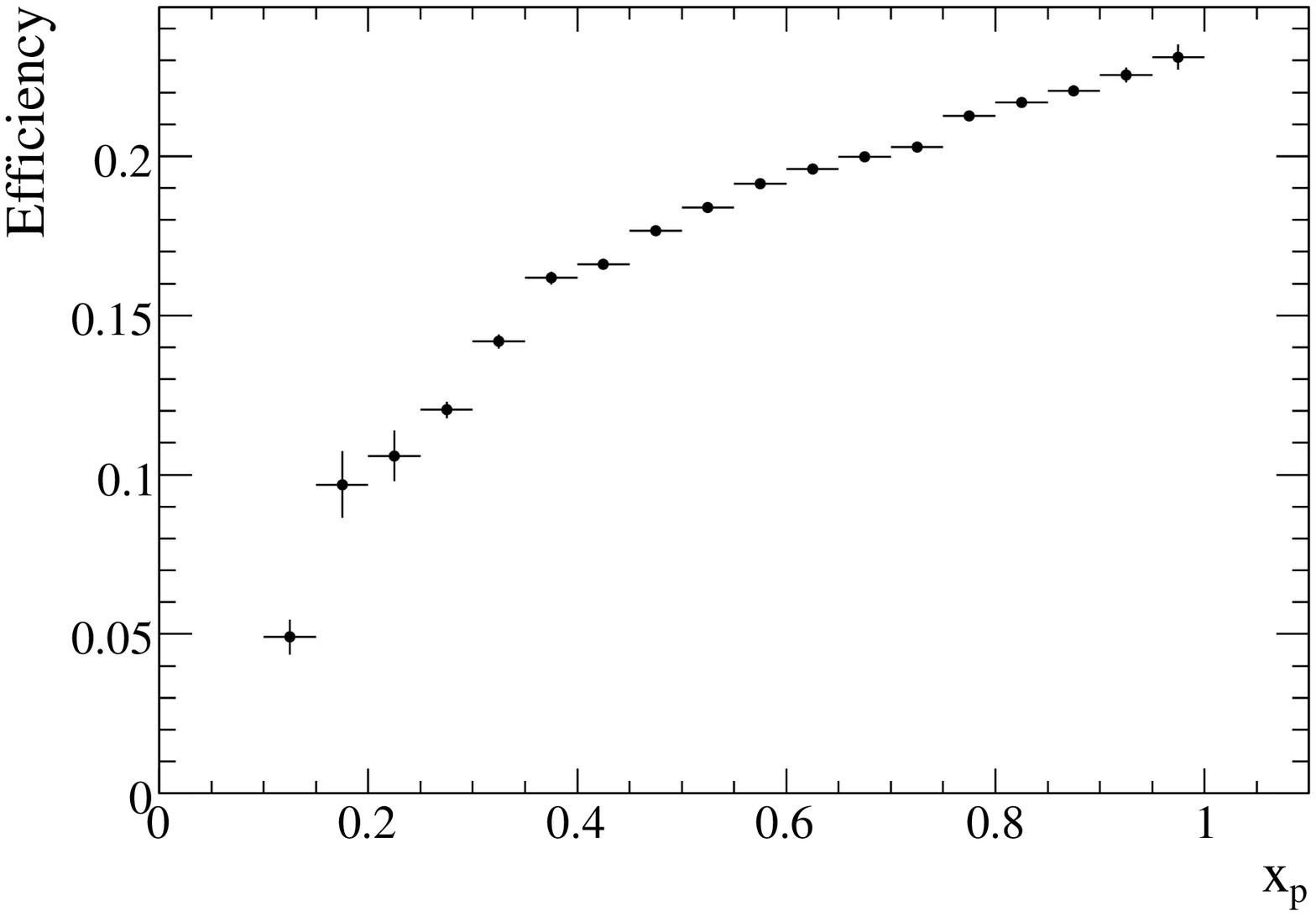}
\enf{fig:eff}{Reconstruction efficiency for the decay chain $\Y2S \to
  \pi^+\pi^-\Y1S$, $\Y1S\to D^{*\pm} X$ as a function of the scaled $D^{*\pm}$
  momentum \xp.}

\section{Results}
Figure~\ref{fig:xpvphocomparison} shows the efficiency-corrected
distribution of the $D^{*\pm}$ yield as a function of \xp.  The
branching fraction for the inclusive decay $\Y1S\to D^{*\pm} X$ in the
\xp{} range $[0.1,1.0]$ is computed from:
\begin{align}
 \BR[\Y1S \to D^{*\pm} X] &  = \frac{n_{\rm sig}}{ k_{\rm DCS}\times  \BR_{\rm decay}\times  
 N_{\Upsilon(1S)}}  \\
& = (2.52 \pm 0.13({\rm stat}) \pm 0.15({\rm syst}) )\% \notag
\end{align}
where $n_{\rm sig}=\Sigma_{\xp}n_{\rm sig}(\xp)/\epsilon(\xp)=11845 \pm 596$ is
the efficiency-corrected signal yield in the \xp{} range $[0.1,1.0]$,
$k_{\rm DCS}=(99.62\pm 0.02)\%$ is a correction factor to account for the
subtraction of DCS $D^0$ decays, $\BR_{\rm decay}$ is the product of the
branching fractions~\cite{pdg08} in the $D^{*\pm}$ decay chain $\BR[D^{*+}\to D^0
  \pi^+] = (67.7\pm 0.5)\%$ and $\BR[D^0 \to K^-\pi^+]=(3.91 \pm
0.05)\%$, $N_{\Upsilon(2S)} = (98.6 \pm 0.9) \times 10^6$, and
$N_{\Upsilon(1S)} = N_{\Upsilon(2S)} \times \BR[\Upsilon(2S) \to
  \pi^+\pi^- \Y1S] = (17.8 \pm 0.4) \times 10^6$ is the number of \Y1S
mesons produced in this decay chain.

\bef \includegraphics[width=0.49\textwidth]{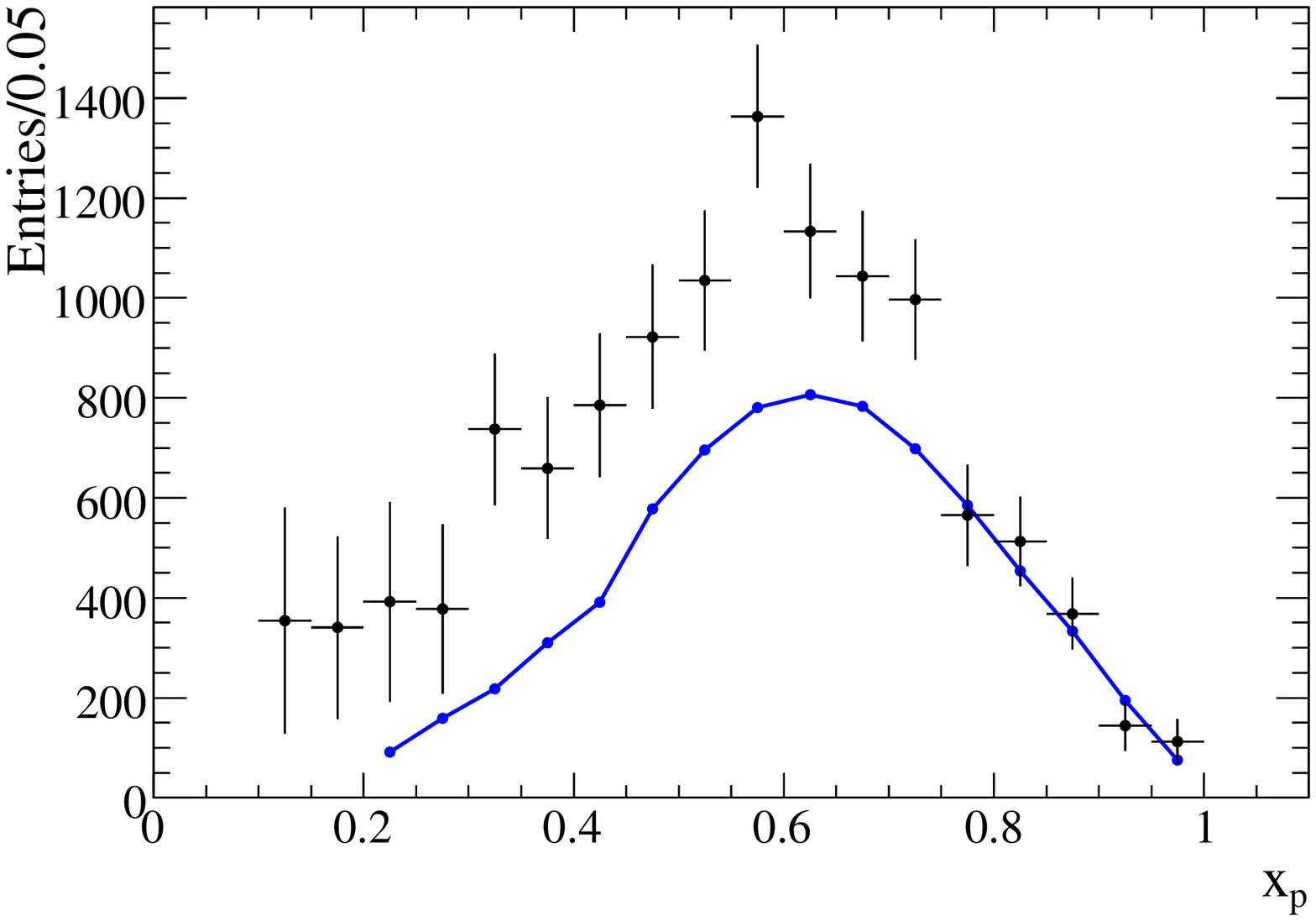}
\enf{fig:xpvphocomparison}{Signal yield as a function of \xp. The
  solid line represents the expected contribution from the virtual
  photon process~\cite{cleo_frag}.}

\bet{lc}{tab:systematic}{Summary of the systematic uncertainties on $\BR[\Y1S \to D^{*\pm} X]$.}
Sources of systematic uncertainty  &  Relative uncertainty  \\
\hline
Slow $\pi^\pm$ reconstruction efficiency  &  3.0\%           \\
$M_{\rm recoil}$ selection                 &  2.8\%           \\
$\BR_{\rm decay}\times \BR[\Upsilon(2S) \to \pi^+\pi^- \Y1S]$ &  2.3\%   \\
Generated \xp{} distribution             &  2.2\%           \\
PID                                      &  1.6\%           \\
Tracking efficiency (excl. slow pion)    &  1.6\%           \\
\Y2S decay model                         &  1.2\%           \\
$\Upsilon$ counting                      &  0.9\%           \\
Background curvature                     &  0.4\%           \\
MC efficiency                            &  0.4\%           \\
Signal shape                             &  0.3\%           \\
$k_{\rm DCS}$                             &  0.02\%          \\
\hline
Total systematic uncertainty             & 5.9\%            \\
\ent

We verify that our analysis procedure is unbiased by fitting
off-resonance data and a Monte Carlo simulation of the background; 
we find no significant signal. We also compare the lower and upper
sidebands and use the $D^0$ mass sidebands instead of the recoil mass
to subtract the background, and we find no significant shift in the
signal. The sources of systematic uncertainty are listed in
Table~\ref{tab:systematic}. The main contributions come from the
uncertainties in the knowledge of the slow pion reconstruction
efficiency and the selection efficiency of \Y1S decays in the recoil
mass signal region.  The former is determined from a control sample of
$D^{*+}\to D^0\pi^{+}$ decays by comparing the efficiency in data with
that in MC events, for the soft pion momentum range $[50,400]$
\mev. The efficiency is extracted from a study of the angular
distribution of the soft pion in the rest frame of the $D^*$ meson.
The $M_{\rm recoil}$ selection systematic uncertainty is obtained by
comparing the recoil mass distribution for signal events in the full
\xp{} range $[0.1,1.0]$ in data, with the distribution in Monte Carlo
simulated events. The fit to data with the sum of two Gauss functions
gives an r.m.s. of $2.9 \mev$ while the fit to MC events gives $3.3
\mev$.  The efficiency is estimated from the integral of the fitted
function in a window around the \Y1S mass of $\pm 2\times$ the
r.m.s. on MC (the recoil mass signal region). The efficiency in data
is 96.3\% while in MC events is 93.6\%, which corresponds to a relative
systematic error on the result of 2.8\%.
The uncertainty associated with
the generated \xp{} distribution is determined by reweighting
simulated signal MC events according to the \xp{} distribution
measured using data. In addition the parameters of the \Y2S decay
model have been varied within their uncertainty and the resulting
relative efficiency variation has been taken as the systematic uncertainty.
The uncertainty in the particle identification efficiency (PID) is
derived from a study of a $\phi\to K^+K^-$ control sample and by
removing the PID requirement from the selection. The dominant
systematic uncertainties in the $\Upsilon$ counting come from the
modeling of the track reconstruction efficiency and of the total
energy of the events. The signal shape uncertainty is due to data-MC
differences in the $D^0$ mass signal distribution. A possible
curvature of the background is extracted from off-resonance data, and
the systematic uncertainty is obtained by adding the corresponding second
order polynomial to the background p.d.f.. The uncertainties due to MC
efficiency, $k_{\rm DCS}$ and $B_{\rm decay} \times \BR[\Upsilon(2S)
  \to \pi^+\pi^- \Y1S]$ arise from imperfect knowledge of these
parameters.

\section{Discussion and Conclusion}

Figure~\ref{fig:xpvphocomparison} shows the expected \xp{}
distribution for $D^{*\pm}$ production from the QED virtual photon
annihilation process, $\Y1S\to\gamma^* \to c\bar c$. The shape
is obtained from the measured $D^{*\pm}$ fragmentation function at
$\sqrt{s}=10.5~\gev$ ~\cite{cleo_frag} and the normalization is
computed from:
\begin{align}
\label{eq:vpho}
\BR[\Y1S \to \gamma^* \to D^{*\pm} X] = 
& \frac{\sigma_{D^{*\pm}}}{ \sigma_{q\bar q}} \times R_{\rm had} \\
& \times \BR[\Y1S \to \mu^+\mu^-] \notag
\label{eq:vpho}
\end{align}
where, $R_{\rm had} = \sigma(e^+e^-\to {\rm hadrons})\slash\sigma(e^+e^- \to
\mu^+\mu^-)=3.46\pm0.13$\cite{argusR}, $\BR[\Y1S \to
  \mu^+\mu^-]=(2.48\pm 0.05)\%$\cite{pdg08}, and
$\frac{\sigma_{D^{*\pm}}}{ \sigma_{\rm q\bar q}}=(17.7\pm
2.2)\%$\cite{pdg08} is the measured $D^{*\pm}$ yield from $e^+e^-\to
q\bar q$ at $\sqrt{s}=10.5~\gev$. We find $\BR[\Y1S \to
  \gamma^* \to D^{*\pm} X]= (1.52 \pm 0.20)\%$ .

Our measured branching fraction exceeds the expected rate from the QED
virtual photon process from Eq.~\ref{eq:vpho} by $(1.00 \pm 0.28)\%$
(including the systematic uncertainty) which corresponds to $3.6$ standard
deviations. While the measured \xp{} spectrum agrees in shape with
that of the virtual photon process for $\xp > 0.75$, there is a
significant excess for $\xp < 0.75$.  The probability that the measured
spectrum is consistent with the expected distribution from the virtual
photon, normalized using Eq.~(\ref{eq:vpho}), is $1.2\times 10^{-5}$ 
confidence estimated from a binned $\chi^2$ test.
The excess is compatible with the contribution
expected~\cite{singlet3} from the splitting of a virtual gluon, $(1.20
\pm 0.29)\%$. This does not leave much room for the octet
contribution~\cite{octet}, which is also disfavored from the shape of
the excess as a function of \xp.

In summary, using the data collected with the \babar{} detector at the
$\Upsilon(2S)$ resonance, we have observed for the first time the
decay of \Y1S mesons to open charm. We have measured the branching
fraction $\BR[\Y1S \to D^{*\pm} X] = (2.52 \pm 0.13({\rm stat}) \pm
0.15({\rm syst}) )\%$ and the $D^{*\pm}$ momentum distribution in the rest
frame of the \Y1S.  We find evidence for a significant excess of
$D^{*\pm}$ production with respect to the expectation from the virtual
photon annihilation process.

\begin{acknowledgments}
We are grateful for the 
extraordinary contributions of our \pep2\ colleagues in
achieving the excellent luminosity and machine conditions
that have made this work possible.
The success of this project also relies critically on the 
expertise and dedication of the computing organizations that 
support \babar.
The collaborating institutions wish to thank 
SLAC for its support and the kind hospitality extended to them. 
This work is supported by the
US Department of Energy
and National Science Foundation, the
Natural Sciences and Engineering Research Council (Canada),
the Commissariat \`a l'Energie Atomique and
Institut National de Physique Nucl\'eaire et de Physique des Particules
(France), the
Bundesministerium f\"ur Bildung und Forschung and
Deutsche Forschungsgemeinschaft
(Germany), the
Istituto Nazionale di Fisica Nucleare (Italy),
the Foundation for Fundamental Research on Matter (The Netherlands),
the Research Council of Norway, the
Ministry of Education and Science of the Russian Federation, 
Ministerio de Educaci\'on y Ciencia (Spain), and the
Science and Technology Facilities Council (United Kingdom).
Individuals have received support from 
the Marie-Curie IEF program (European Union) and
the A. P. Sloan Foundation.

\end{acknowledgments}

\end{document}